\documentstyle[11pt,epsfig,aaspp4]{article}
\begin{document}

\title{The Velocity Dispersion of MS1054--03: A Massive Galaxy
Cluster at High Redshift$^1$}
\altaffiltext{1}{Based on observations obtained at the W. M. Keck
Observatory, which is operated jointly by the California Institute of
Technology and the University of California.}

\author{
Kim-Vy H. Tran \altaffilmark{2}\altaffiltext{2}{Department of Astronomy and
Astrophysics, University of California, Santa Cruz, CA 95064;
vy@ucolick.org},
Daniel D. Kelson \altaffilmark{3}\altaffiltext{3}{Department of
Terrestrial Magnetism,
Carnegie Institution of Washington, 5241 Broad Branch Road, NW,
Washington D.C., 20015; kelson@gimpel.ciw.edu},
Pieter van Dokkum \altaffilmark{4,5}\altaffiltext{4}{Leiden Observatory, 
P.O. Box 9513, 2300 RA Leiden, The Netherlands; 
dokkum@astro.rug.nl, franx@astro.rug.nl}
\altaffiltext{5}{Kapteyn Astronomical Institute,
P.0. Box 800, NL-9700 AV, Groningen, The Netherlands},
Marijn Franx \altaffilmark{4,5},
Garth~D.~Illingworth \altaffilmark{6}\altaffiltext{6}{University of California
Observatories/Lick Observatory, University of California, Santa Cruz,
CA 95064; gdi@ucolick.org, magee@ucolick.org}, 
and Daniel Magee \altaffilmark{6} }

\begin{abstract}

We present results from a dynamical study of the high redshift,
massive, X-ray luminous galaxy cluster MS1054--03.  We significantly
increase the number of confirmed cluster members by adding 20 to an
existing set of twelve; using the confirmed members, we estimate
MS1054--03's redshift, velocity dispersion, and mass.  We find that
$z=0.8329 \pm 0.0017$, $\sigma = 1170 \pm 150$ km/s, and the central
mass is approximately $1.9 \pm 0.5 \times10^{15} h^{-1} M_{\odot}$
(within $R=1 h^{-1}$ Mpc; $H_0 =100h$ km s$^{-1}$ Mpc$^{-1}$,
$q_0=0.5$).  MS1054--03 is one of a handful of high redshift ($z>0.5$)
clusters known that also has X-ray and weak-lensing observations
(Donahue et al. 1998; Luppino \& Kaiser 1997); we find our dynamical
mass agrees with mass estimates from both studies.  The confirmation
of MS1054--03 as a massive cluster at $z\sim0.8$ is consistent with an
open ($\Omega_M\sim0.3$) or flat, $\Lambda$-dominated
($\Omega_M+\Omega_{\Lambda}=1$) universe.  In addition, we compare
MS1054--03's velocity dispersion and X-ray temperature to a sample of
low and intermediate redshift galaxy clusters to test for evolution in the
$\sigma - T_x$ relation; we find no evidence for evolution in this
relation to $z\sim0.8$.

\end{abstract}

  
\section{Introduction}

With the advent of large telescopes and X-ray surveys (e.g. {\it
Einstein} Medium Sensitivity Survey, ROSAT North Ecliptic Pole
Survey), the study of high redshift ($z>0.5$) galaxy clusters has
evolved into a field rich with multi-wavelength observations (Gioia \&
Luppino 1994; Fukazawa et al. 1994; Carlberg et al. 1996; Donahue
1996; Donahue et al. 1998).  Forming at the junction of walls and
filaments (Kang et al. 1994; Cen \& Ostriker 1994), clusters represent
the extreme end of fluctuations in the primordial power spectrum and
may place strong constraints on cosmological models (Eke et al. 1998;
Gioia 1998).  On a smaller scale, the cluster environment provides an
useful laboratory for studying galaxy evolution in a range of
different local densities.  While low redshift clusters have been
studied for over half a century now, the discovery of high redshift
galaxy clusters (Gioia \& Luppino 1994; Gioia 1998) has opened a new
avenue for using them as tools to probe the evolution of large scale
structure and galaxies from $z \sim 1$ to the present.

The existence of massive clusters at high redshift may constrain the
mean matter density of the universe ($\Omega_M$).  In a high density
universe ($\Omega_M \sim 1$), massive clusters would have formed
fairly recently and their main epoch of growth would be from low
redshift ($z\sim0.3$) to the present (Carlberg et al. 1997).  In this
model, the existence of massive clusters at redshifts greater than 0.5
is highly unlikely and their number density evolves quickly with
redshift (Gross et al. 1998; Carlberg et al. 1997).  In a low density
universe, however, structure formed early and quickly, ``freezing
out'' at higher redshift and so the number density evolution is much
milder.  A flat, $\Lambda$-dominated ($\Omega_M+\Omega_{\Lambda}=1$)
universe predicts slightly stronger evolution than a open, low
$\Omega_M$ model but the results are similar (Bahcall, Fan, \& Cen
1997).  The difference in the predicted number of massive clusters at
$z\sim0.8$ between low and high density models is several orders of
magnitude; Bahcall \& Fan (1998) and Donahue et al. (1998) quote a
factor of $\sim 10^5$.  As such, the existence of a few massive
($10^{14} M_{\odot}$ ), high redshift galaxy clusters can rule out a
high $\Omega_M$ universe (Gioia 1998; Bahcall \& Fan 1998; Gross et
al. 1998).

Presently, there are three favored methods to measure cluster masses:
measuring the cluster velocity dispersion (pioneered by Zwicky 1933);
mapping the X-ray emissivity of the intracluster gas (Cavaliere \&
Fusco-Femiano 1976; Cavaliere \& Fusco-Femiano 1978); or using weak
and/or strong-lensing to trace the cluster mass distribution
(Bartelmann 1995; Miralda-Escud\'e \& Babul 1995; Fort \& Mellier
1994).  Each method, however, has uncertainties resulting from
different sources which may over- or underestimate the mass
significantly.  For example, lensing traces the total matter
distribution in a cluster but a cluster's weak-lensing map is affected
by any additional mass along the line of sight from the observer to
the galaxies serving as the background sources, and the redshift
distribution of the background sources is a considerable source
of error.  Lensing is also affected by the flat-sheet dilemma
(Bartelmann 1995) which causes one to underestimate the true mass.
As for using the velocity dispersion or X-ray emissivity,
these are easily affected by cluster substructure; accretion of
sub-groups can increase the former (Crone \& Geller 1995), and
deviation of the intracluster gas from hydrostatic equilibrium can
introduce errors up to 50\% in the mass estimate (Roettiger, Burns, \&
Loken 1996).  To overcome the uncertainties inherent in each method,
it is best to use a combination of all three to study a cluster's 
dynamics (Markevitch 1998; Smail 1997).

In this paper, we weigh the high redshift galaxy cluster MS1054--03 by
measuring its velocity dispersion.  This work complements the
weak-lensing study completed by Luppino \& Kaiser (1997; hereafter
LK97) and the X-ray study by Donahue et al. (1998; hereafter D98) of
this cluster.  Our results are based on spectra collected with the
Keck~II Telescope of 24 cluster members.  With this sample, we measure
MS1054--03's velocity dispersion ($\sigma$), estimate the
corresponding dynamical mass, and compare our results to X-ray (D98)
and weak-lensing (LK97) results.  In addition, it has been observed
that there is no evolution in the $\sigma - T_X$ relation for a sample
of lower redshift clusters ($z\leq0.54$; Mushotsky \& Scharf
1997; hereafter MS97); we place MS1054--03 on the $\sigma - T_X$ plane
to test this result at high redshift.

In our calculations, we use $H_0 = 100 h$ km s$^{-1}$ Mpc$^{-1}$,
$q_0=0.5$, and $\Lambda=0$ except where noted.

\section{Data}

Our objects were chosen from two target lists.  Higher priority
objects were cluster member candidates selected based on their $I$
fluxes ($I<22.1$) and two colors, $(R-I)$ and $(B-R)$.  Lower priority
objects were faint field and additional cluster member candidates
selected based on their $I$ magnitude alone ($22<I<24$).  Our main
criterion was to identify cluster members bright enough to measure
absorption line velocity dispersions using the G-band (van Dokkum et
al. 1998); the fainter galaxies were assigned lower priority on the
slit-masks.  The $B$ and $R$ images were kindly made available by
G. Luppino and are described in LK97; the $I$ image was taken with
Keck/LRIS.  The package FOCAS ({\it Faint Object Classification and
Analysis System}; Valdes 1982) was used to measure the fluxes (total
light within $r\sim 1.''2$).  The target selection did not include
galaxy morphology.

The spectra were taken with the Keck~II Telescope in February 1997
during a two night run.  Four multi-slit masks were used to cover a
$6'\times7.'8$ field; at MS1054--03's redshift, this field corresponds
to a region approximately $1.5 \times 1.9 h^{-1}$ Mpc.  Using the {\it
Low Resolution Imaging Spectrograph} (LRIS; Oke et al. 1995) with the
831 mm$^{-1}$ grating centered at 8200 \AA~(instrumental resolution
$\sigma_{instr}\sim 50$ km/s), we integrated for two hours each on
three of the masks and 2.6 hours on the fourth.  Of the original 110
targets on the masks, useful spectra were obtained for 52 objects; the
lost spectra were due to low signal to noise, scattered light, or a
combination of both.  A bright blue star also was included on all four
masks to correct for the H$_2$O atmospheric absorption feature (7600
\AA).  The seeing was $\sim 1''$ on both nights.

A combination of IRAF packages and customized programs were used to
reduce the multi-slit spectra.  The spectra were cleaned of cosmic
rays using software made available by A. Phillips; the spectra were
then flat-fielded, rectified, and wavelength calibrated using the
software package Expector (Kelson 1998).  The sky subtraction and the
extraction of the spectra were done in the usual way using IRAF.  The
extracted spectra cover a wavelength range of $\sim 6800 - 9400$ \AA~
with a dispersion of $\sim$~1~\AA/pixel.

To determine redshifts, we used the IRAF task XCSAO (Kurtz et
al. 1992) to cross-correlate the 52 spectra with four galaxy
templates: NGC7331 (morphological type SA(s)b), NGC4889 (E4), NGC2276
(SAB(rs)c), and an E+A galaxy.  The E+A spectrum was created by adding
an A star spectrum to NGC4889.  In our wavelength range, the main
features in a cluster galaxy spectrum are H$\delta$, CaI (4227 \AA),
G-band, H$\gamma$, Fe (4383.6 \AA), and H$\beta$; in some members, the
H and K break also is visible.  We found that NGC7331 and the E+A
galaxy were the best templates to use in confirming cluster members.  The
24 galaxies confirmed to be cluster members are listed in Table 1,
along with their heliocentric redshifts and positions.

%\placetable{table1}

In the same table, we include the eight additional cluster members
from the D98 study.  D98 actually has spectra for 12 cluster members
but four overlap with our sample; we use our redshifts for these four
since our errors are smaller.  Comparison of the four overlapping
redshifts shows that ours are slightly higher ($\overline{\delta z}
\approx 0.0024\pm0.0044$).  Although these differences are within
$1\sigma$ of the estimated errors, we choose not to include D98's
redshifts in our final analysis since the offset in the four common
members may indicate a slight bias between the two data sets.

The redshift errors for the galaxies in our set are small.  Since we
intended to measure dispersions of individual cluster members (van
Dokkum et al. 1998), the spectra have unusually high signal-to-noise
for a redshift survey which results in small errors.  A combination of
the grating's high spectral resolution ($\sigma_{instr}\sim 50$km/s),
the large number of sky lines used in the wavelength calibration, and
the multiple absorption features used in the cross-correlation routine
also reduced the errors.  The dominant factor in the redshift error is
the instrumental resolution of LRIS.

\section{Results}

In Fig. 1, we present an $I$ image of the field with the 32 confirmed
cluster members marked (including the eight from D98); galaxy 1484 is
the Brightest Cluster Galaxy (BCG).  The image is approximately $5.'1$
on a side ($1.3 h^{-1}$ Mpc at $z=0.83$).  MS1054--03's striking
structure is seen as the filament stretching from east to west; there
also appears to be structure north and south of the cluster core.

%\placefigure{imI}

The velocity distribution for the field and cluster galaxies is
plotted in the upper panel of Fig. 2.  Note how MS1054--03 stands out
as a strong peak at $z=0.83$.  The lower panel of Fig. 2 contains a
histogram of the 24 cluster members (noted by the solid boxes).  The
bin size (200 km/s) corresponds to three times the average error in
the individual galaxy redshifts.  In the same panel, D98's 12 galaxies
(dotted boxes) are also included and the bin size adjusted to their
errors.

%\placefigure{hist}

To determine the mean redshift and velocity dispersion of MS1054--03,
we use the biweight, bootstrap, and jacknife methods of Beers, Flynn,
\& Gebhardt (1990) since they have proven to be robust estimators when
dealing with small samples ($N<50$).  The biweight estimator is used
to measure both the cluster's redshift and its velocity dispersion.
The corresponding errors are estimated using the bootstrap (redshift)
and jacknife (velocity dispersion) algorithms.  All of these methods
methods take into account the associated error in the measurements.
None of these methods assume that the cluster member velocity
distribution is Gaussian.

Using the 24 confirmed members from our sample, we measure the cluster
redshift to be $z=0.8329 \pm 0.0017$, and the velocity dispersion to
be $\sigma = 1170\pm 150$ km/s; the latter is corrected to the cluster
rest-frame by dividing by the factor $(1+z)$ (Peebles 1993).  If we
include D98's eight members in our weighted analysis and correct them
for the systematic offset of $\overline{\delta z} \approx 0.0024 \pm
0.0044$ , the cluster's redshift decreases slightly ($0.8323 \pm
0.0017$) and the velocity dispersion increases to $1230 \pm 140$ km/s.
Due to the offset between our sample and D98's, however, we use only
our 24 members in the following analysis.  Like Carlberg et
al. (1996), we find that with more cluster members (24), the velocity
dispersion decreases from the previous estimate which used only 12
members ($\sigma_{D98} =1360\pm450$).

To estimate the mass using the velocity dispersion, we follow Ramella,
Geller, \& Huchra (1989) (also Nolthenius \& White 1987) by first
determining the cluster's virial radius:
\begin{equation}
R_V = \frac{\pi \bar{z}}{H_0} \left\{ \frac{1}{2} \left[
\frac{N_{mem}(N_{mem}-1)}{2} \left( \sum_i \sum_{j>i} \theta^{-1}_{ij}
\right)^{-1}\right]\right\} 
\end{equation}
where $\bar{z}$ is the redshift of the cluster, $N_{mem}$ is the number of
cluster members, and $\theta_{ij}$ is the angular separation of
cluster members $i$ and $j$.  The cluster's virial mass follows as
\begin{equation}
M = \frac{6\sigma^2_{1D} R_V}{G}
\end{equation}
where $M$ is the mass, $\sigma_{1D}$ is the line-of-sight velocity
dispersion ($\sigma = 1170\pm 150$ km/s), and $R_V$ is the virial
radius.  We determine $R_V$ to be $1 h^{-1}$ Mpc, and the
corresponding mass to be $M = 1.9 \times10^{15} h^{-1} M_{\odot}$.
Using the error in the velocity dispersion, the corresponding error in
our mass estimate is approximately $0.5\times 10^{15} h^{-1}
M_{\odot}$ ($\sim25\%$).

We note that our simple method of estimating the mass does not take
into account systematic errors which easily can change the mass
estimate by a factor of two (Crone \& Geller 1995; Cen 1996).  Like
many clusters, MS1054--03 is elongated along the plane of the sky (de
Theije, Katgert, \& van Kampen 1995; Binggeli 1982) with the main
structure extending from east to west (see Fig. 1).  The same
elongation is seen in the X-ray and weak-lensing maps, so MS1054--03
may not be virialized or it may be triaxial, or it may be both.  A
dynamical treatment such as this is sensitive to non-virialization,
deviation from an isothermal profile, substructure, and triaxiality;
X-ray and lensing estimates also are sensitive to these factors but to
different degrees.  Thus, the formal errors quoted by the three
methods used to estimate MS1054--03's mass may be overshadowed by the
errors introduced by these effects.

\section{Discussion}

\subsection{Comparison to X-ray and Weak-Lensing Results}

D98 have measured MS1054--03's X-ray temperature with ASCA and mapped
the luminosity of the cluster's intracluster medium with the ROSAT
HRI.  By adopting an isothermal model in a matter-dominated universe
($\Omega_M=1$), they use the X-ray temperature ($12.3\pm^{3.1}_{2.2}$
keV) to estimate the cluster's virial mass; the virial mass is chosen
to correspond to a volume where the mean density is 200 times the
critical density.  Within this characteristic radius ($r_{200}=1.5
h^{-1}$ Mpc), the estimated mass is $0.74 \times 10^{15} h^{-1}
M_{\odot}$.  The difference in the X-ray and dynamical mass estimates
may be due to difficulties in determining the correct shape,
characteristic radius, and mass distribution of any cluster.  For
example, projection effects and nonequilibrium of the intracluster gas
with the potential can result in an underestimate of the X-ray
temperature and introduce errors up to 50\% in the mass (Roettiger,
Burns, \& Loken 1996).  D98 also note that measuring the virial masses
of clusters becomes more difficult then measuring their X-ray
temperatures with increasing redshift since virial masses depend on
the adopted cosmology.  While the two mass estimates differ, however,
they do agree within their large uncertainties and both do support the
main result which is that MS1054--03 is a massive cluster.

For the weak-lensing analysis, LK97 use ground-based images of
MS1054--03 to estimate its mass distribution ($H_0=100h$ km s$^{-1}$
Mpc$^{-1}$, $\Omega_M=1$) out to a radius of $1 h^{-1}$ Mpc.  In their
models, the cluster's total enclosed mass depends on the redshift of
the background sources ($z_s$); since this is unknown, they consider
models where the background galaxies lie in sheets at $z_s= 1, 1.5,
\&~3$.  Depending on whether $z_s$ is 3 or 1, the mass within a radius
of $1 h^{-1}$ Mpc can differ by more than a factor of five, $1\times
10^{15} h^{-1} M_{\odot}$ to $5\times10^{15} h^{-1} M_{\odot}$
respectively.  We find our mass estimate ($1.9\pm0.5\times10^{15}
h^{-1} M_{\odot}$) best agrees with a weak-lensing model where the
sources are at $z_s \sim 3$ if $\Omega_M=1$.  In a low density or
$\Lambda$-dominated universe, however, the redshifts of the background
sources for a given weak-lensing mass estimate will decrease for a
given mass, e.g. from $z_s\sim3$ to $z_s\sim2$ for $M=1\times 10^{15}
h^{-1} M_{\odot}~(R<1 h^{-1}$ Mpc).

The consistency between the three mass estimates for MS1054--03
confirms the existence of at least one massive galaxy
cluster at high redshift ($z>0.5$).  With its high velocity dispersion
and mass, MS1054--03 presents a substantial argument against a flat,
matter-dominated ($\Omega_M=1$) universe (D98; Gross et al. 1998;
Bahcall 1998; Gioia 1998).  In an $\Omega_M=1$ universe, the number
density of clusters evolves strongly from a redshift of 1 to the
present whereas in an open ($\Omega_M \sim 0.3$) or 
$\Lambda$-dominated model, structure forms at higher redshift and the
bulk of clusters are in place by $z\sim1$ (Bartelmann, Ehlers, \&
Schneider 1993).  At $z\sim0.8$, the difference in the predicted
number of clusters between $\Omega_M=1$ and open (or 
$\Lambda$-dominated) models is several orders of magnitudes (Bahcall
\& Fan 1998 and D98 quote a factor of $\sim10^5$), so the likelihood of
finding a high redshift cluster is much greater in an open (or
$\Lambda$-dominated) universe.  Thus, the existence of a handful of
clusters like MS1054--03 may be enough to rule out an $\Omega_M=1$
universe (Gross et al. 1998; Carlberg 1997).

\subsection{$\sigma-T_X$ Relation}

In Fig. 3, we follow earlier work by MS97 at lower redshift by comparing
MS1054--03's velocity dispersion and X-ray temperature to values
measured for other clusters ($0.19<z<0.55$).  In Table 2, we list the
clusters, their velocity dispersions, X-ray temperatures, and
references.  These particular clusters were selected from the literature
based on their redshifts and that their X-ray temperatures were
measured using the ASCA satellite.  In Fig. 3, we fit a curve to the
data using weighted least squares.  The form of the fit is
$T_X=a\sigma^n$ where both the constant $a$ and the power $n$ are
variables; the fit is weighted by the errors in both $\sigma$ and
$T_X$.  As previous workers have done (Edge \& Stewart 1991; MS97;
D98), we include a curve denoting the virial relation $kT_X\beta=\mu
m_p \sigma^2$ with $\mu=0.6$ and $\beta=1.0$.

%\placetable{table2}
%\placefigure{xray}

Despite being the hottest ($T_X=12.3 \pm^{3.1}_{2.2}$ keV) and most
distant cluster in the sample, MS1054--03 lies on the same trend as
the lower redshift clusters, a result which suggests little or no
evolution in the $\sigma-T_X$ relation.  Also interesting is that the
$\sigma-T_X$ relation for these clusters, including MS1054--03, follow
the virialized model fairly well, indicating that both the X-ray gas
and galaxies trace the same gravitational potential well.  This has
been noted by MS97 for a sample of lower redshift clusters
($0.14<z<0.55$).  Our current work, which includes significantly more
clusters than D98 in the redshift range $0.19<z<0.83$, confirms MS97's
conclusions to $z\sim0.8$.  Since MS1054--03 does not appear to be
substantially less evolved than its lower redshift counterparts, it
further suggests a low $\Omega_M$ or $\Lambda$-dominated universe
since in these models cluster structure does not evolve significantly
from $z\sim0.8$ to now (Bartelmann et al. 1998; Bartelmann, Ehlers, \&
Schneider 1993).

An argument against MS1054--03 being as evolved as low redshift
clusters is its elongation along the plane of the sky.  It should be
noted, however, that such structure is seen in some low and
intermediate redshift clusters (White et al. 1993; Bird, Davis, \&
Beers 1995; Markevitch et al. 1998) and may indicate triaxiality
rather than non-virialization.  A further investigation of triaxility
and substructure is not possible with the present set of 32 members.
MS1054--03's agreement with trends relating X-ray temperatures and
velocity dispersions derived from low and intermediate redshift
clusters (Fig. 3; D98; MS97) suggests that despite its
asphericity and substructure, MS1054--03 may be just as evolved as
these clusters.

\section{Conclusions}

We present a dynamical study of the high redshift galaxy cluster
MS1054--03 using 24 confirmed cluster members ($6'\times7.'8$ field)
to improve D98's estimate of the cluster redshift and velocity
dispersion.  With the 24 members, we find that MS1054--03 has a mean 
$z=0.8329\pm 0.0017$ and a velocity dispersion of $1170 \pm 150$
km/s.  Its corresponding dynamical mass within $1h^{-1}$ Mpc is
$1.9\pm 0.5\times10^{15} h^{-1} M_{\odot}$.

We find that the dynamical and X-ray mass estimates agree within the
errors, leading us to conclude that the intra-cluster gas and galaxies
may be in equilibrium with the cluster's potential (at least to a
radius of $1 h^{-1}$ Mpc).  In addition, comparison of these two mass
estimates with the weak-lensing results places a constraint on the
redshifts of the background sources lensed by MS1054--03; the best
agreement is for $z_s \sim 3$ ($q_0=0.5$).  For this weak-lensing mass
($1\times10^{15} h^{-1} M_{\odot}$), Luppino \& Kaiser (1997) estimate
the corresponding cluster $M/L_V$ to be 350$h~(M/L_V)_{\odot}$.

With this velocity dispersion and an X-ray temperature of $12.3
\pm^{3.1}_{2.2}$ keV (D98), MS1054--03 lies on the same trend in the
$\sigma - T_X$ relation as a sample of lower redshift clusters
($0.19<z<0.55$).  This consistency between MS1054--03 and the lower
redshift sample supports no evolution in the $\sigma - T_X$ relation
to $z\sim0.8$.  In addition, the agreement of these clusters with the
virial relation $kT_X\beta=\mu m_p \sigma^2$ (with $\beta=1.0$ and
$\mu=0.6$) is consistent with both the X-ray gas and galaxies tracing
the same gravitational well even at high redshift.

Despite MS1054--03's high redshift and aspherical morphology, the
consistency between our results with X-ray and weak-lensing studies
argues for a well-developed cluster core similar to those at lower
redshift.  Certainly, there is little disputing MS1054--03's mass, a
result which is difficult to accommodate in a high $\Omega_M$
universe.  The lack of evolution in the $\sigma - T_X$ relation to a
redshift of $z\sim0.8$ also argues for early structure formation and
thus a low density or $\Lambda$-dominated model.  Although these
results do not effectively rule out a high density universe, they do
add to the mounting support for a low density (or $\Lambda$-dominated)
one.

In the future, we plan to continue our dynamical study of MS1054--03
by adding more cluster members to our present set; the larger set will
allow us to probe the cluster's substructure and refine our naive
approach of assuming spherical symmetry and hydrostatic equilibrium to
measure the mass.  We will combine the spectra with an HST
WFPC2 mosaic of the cluster (van Dokkum, in preparation) taken in May
1998.  With the spectra and high resolution images, we will probe
MS1054--03's optical substructure, examine the individual galaxy
profiles of cluster members, and better compare MS1054--03 to galaxy
clusters at lower redshift.

\acknowledgements

We would like to acknowledge G. Luppino, I. Gioia, and M. Donahue for
the generous use of their data.  We thank A. Zabludoff for her
insightful thoughts on this paper, and the anonymous referee for his
comments.  We also appreciate the efforts of those at the Keck
telescope who developed and supported the facility and instruments
that made this program possible.  Support from STScI grants
GO-05991.01-94A, AR-05789.01-94A, and GO-07372.01-96A is gratefully
acknowledged.

\newpage

\clearpage

\makeatletter
\def\jnl@aj{AJ}
\ifx\revtex@jnl\jnl@aj\let\tablebreak=\nl\fi
\makeatother

\begin{deluxetable}{lrrr}
\tablewidth{0pc}
\tablecaption{Redshifts of Cluster Members \label{table1}}
\tablehead{
\colhead{Galaxy $\#$} 	& \colhead{$z$} & 
\colhead{Offset E/W\tablenotemark{a}}& 
\colhead{Offset N/S\tablenotemark{a}} \\
\colhead{} 	& \colhead{} &
\colhead{($+/-$ arcsec)} & \colhead{($+/-$ arcsec)}	}
\startdata
0696 & 0.8312 $\pm$ 0.0002 & 58 & -84  \nl
0997 & 0.8390 $\pm$ 0.0002 & -136 & -58  \nl
1163 & 0.8329 $\pm$ 0.0002 & 63 & -29  \nl
1209 & 0.8380 $\pm$ 0.0002 & -84 & -24  \nl
1280 & 0.8371 $\pm$ 0.0002 & -122 &  -17  \nl
1294 (D5) & 0.8353 $\pm$ 0.0002 & -22 & -13  \nl
1325 & 0.8317 $\pm$ 0.0002 & -53 & -10 \nl
1329 (D2) & 0.8346 $\pm$ 0.0002 & 24 & -6  \nl
1340 & 0.8403 $\pm$ 0.0002 & -45 & -2  \nl
1359 (D10) & 0.8175 $\pm$ 0.0002 & -39 & 0  \nl
1405 & 0.8367 $\pm$ 0.0002 & 46 & -4  \nl
1430 & 0.8239 $\pm$ 0.0002 & 26 & 7  \nl
1457 & 0.8420 $\pm$ 0.0002 & 17 & 0  \nl
1459 & 0.8454 $\pm$ 0.0002 & -6 & 8  \nl
1484 (BCG; D1) & 0.8314 $\pm$ 0.0002 & 0 & 0  \nl
1567 & 0.8282 $\pm$ 0.0002 & 71 & 25  \nl
1583 & 0.8259 $\pm$ 0.0002 & 52 & 23  \nl
1655 & 0.8397 $\pm$ 0.0002 & 38 & 34  \nl
1656 & 0.8224 $\pm$ 0.0002 & 38 & 31 \nl
1701 & 0.8314 $\pm$ 0.0002 & 44 & 48  \nl
1760 & 0.8249 $\pm$ 0.0002 & 34 & 56  \nl
1834 & 0.8392 $\pm$ 0.0002 & 58 & 73  \nl
1942 & 0.8308 $\pm$ 0.0002 & 59 & 98  \nl
1986 & 0.8250 $\pm$ 0.0002 &    134 & 111  \nl
D3 & 0.8127 $\pm$ 0.0003 & 31 & -19  \nl
D4 & 0.8213 $\pm$ 0.0007 & 21 & 21  \nl
D6 & 0.8209 $\pm$ 0.0010 & -29 & -14  \nl
D7 & 0.8286 $\pm$ 0.0010 & -32 & -12  \nl
D8 & 0.8353 $\pm$ 0.0006 & -38 & -8  \nl
D9 & 0.8332 $\pm$ 0.0010 & -44 & -6  \nl
D11 & 0.8378 $\pm$ 0.0030 & -82 & -45 \nl
D12 & 0.8319 $\pm$ 0.0020 & -99 & -39  \nl
\tablenotetext{a}{The offset is given from the central BCG; its
coordinates as measured from an HST image (D98) are $(\alpha, \delta)_{2000} = (10^h 56^m 59.9^s, -3^{\circ}
37' 37.3'')$.} 
\tablecomments{The last eight galaxies in this table are from D98.}
\enddata
\end{deluxetable}

\clearpage
\begin{deluxetable}{lrrcrc}
\tablewidth{0pc}
\tablecaption{Cluster Sample:  Velocity Dispersions and X-Ray Temperatures \label{table2}}
\tablehead{
\colhead{Galaxy Cluster} 	& \colhead{$z$} 	& 
\colhead{$\sigma$ }		& \colhead{Reference ($\sigma$)} 	&
\colhead{$T_X$} 		& \colhead{Reference ($T_X$)}     }
\startdata
A2390 	& 0.2279	& 1093 $\pm$ 61	& 1	& 8.9 $\pm$ 0.9	& 2 \nl
MS0440 	& 0.1965	& 606 $\pm$ 62	& 1	& 5.3 $\pm$ 1.3	& 2 \nl
MS0451+2 (A520) & 0.2010	& 988 $\pm$ 76	& 1	& 8.6 $\pm$ 0.9	& 2 \nl
MS0839 	& 0.1928	& 749 $\pm$ 104	& 1	& 3.8 $\pm$ 0.4	& 3 \nl
MS1008 	& 0.3062	& 1054 $\pm$ 107& 1	& 7.9 $\pm$ 1.2	& 4 \nl
MS1224 	& 0.3255	& 802 $\pm$ 90	& 1	& 4.3 $\pm$ 0.7	& 4 \nl
MS1358 	& 0.3290	& 937 $\pm$ 54	& 1	& 6.6 $\pm$ 0.5	& 4 \nl
MS1455 	& 0.2570	& 1133 $\pm$ 140& 1	& 5.2 $\pm$ 2.2	& 5 \nl
MS1512 	& 0.3726	& 690 $\pm$ 96	& 1	& 3.8 $\pm$ 0.4	& 4 \nl
MS0016 	& 0.5466	& 1234 $\pm$ 128& 1	& 7.6 $\pm$ 0.7	& 6 \nl
MS0451-3& 0.5392	& 1371 $\pm$ 105& 1	& 10.4 $\pm$ 1.2& 7 \nl
MS1054 	& 0.8329	& 1170 $\pm$ 160& This Paper & 12.3 $\pm$ 3.1 & 8 \nl
\tablerefs{(1) Carlberg et al. 1996;  (2) Mushotzky \& Scharf 1997;
(3) Tsuru et al. 1996;  (4) Henry 1997;  (5) Allen et al. 1996;  (6)
Hughes \& Birkinshaw 1995; (7) Donahue 1996;  (8) D98}
\enddata
\end{deluxetable}
\clearpage

\centerline{FIGURE CAPTIONS}

\figcaption[clipI.eps]{A 900 second exposure $I$ image of the field taken with
the Keck I telescope; the field is approximately $5.'1$ on a side
($1.3 h^{-1}$ Mpc at $z=0.83$).  The 24 confirmed cluster members are
shown; the numbers correspond to our galaxy catalog.  Galaxy 1484 is
the brightest cluster galaxy (BCG).  We also identify the eight cluster
members from D98 which did not overlap with our set (see Table
1). \label{imI}}

\figcaption[hist.eps]{{\it Upper Panel:} Histogram (bin size 1000 km/s) of the
60 galaxies for which we have redshifts.  MS1054--03 corresponds to the
strong peak at $z=0.83$.  {\it Lower Panel:} Distribution of the 24 
cluster members in velocity
space (bin size 200 km/s).  The solid line corresponds to our 
data set while the dotted refers to the 12 cluster members from D98
(four galaxies overlap between D98's and our data sets). \label{hist}}

\figcaption[xray.eps]{Comparison of the velocity dispersion of
MS1054--03 ($\sigma$ = 1170 $\pm$ 150 km/s) and X-ray temperature
($12.3 \pm^{3.1}_{2.2}$ keV; D98) to that of 11 clusters at
intermediate redshift ($0.19<z<0.55$); MS1054--03 is the filled
circle.  We fit a curve to the data using a
weighted least squares (solid line) and include the fit's
$1\sigma~rms$ (dotted lines).  The fit is of the form $T_X=a\sigma^n$
where both the constant $a$ and the power $n$ are variables; the fit
is weighted by errors in both $\sigma$ and $T_X$.  Note that the axes
are linear.  Also included is a dashed line denoting the virial
relation $kT_X\beta=\mu m_p \sigma^2$ where $\mu=0.6$ and $\beta=1$.
The least squares fit of $T_X=a\sigma^n$ seems to agree well with a
virialized model for clusters in this redshift range. \label{xray}}

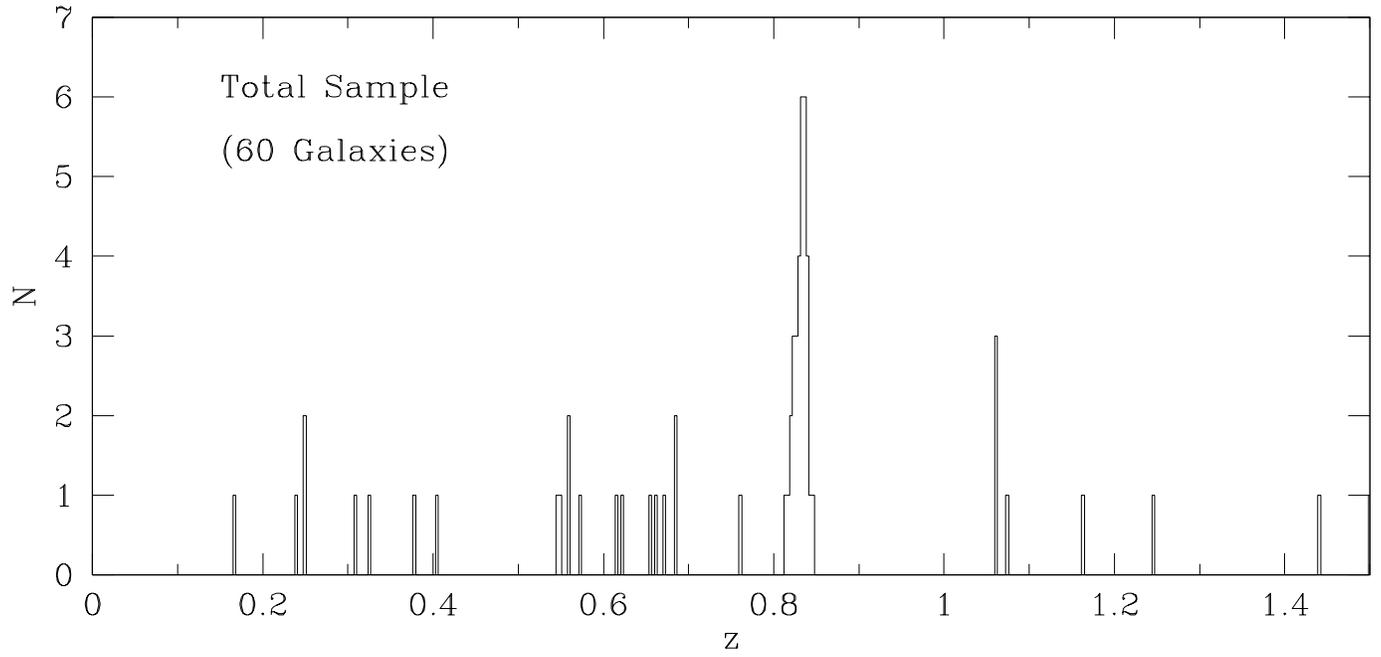
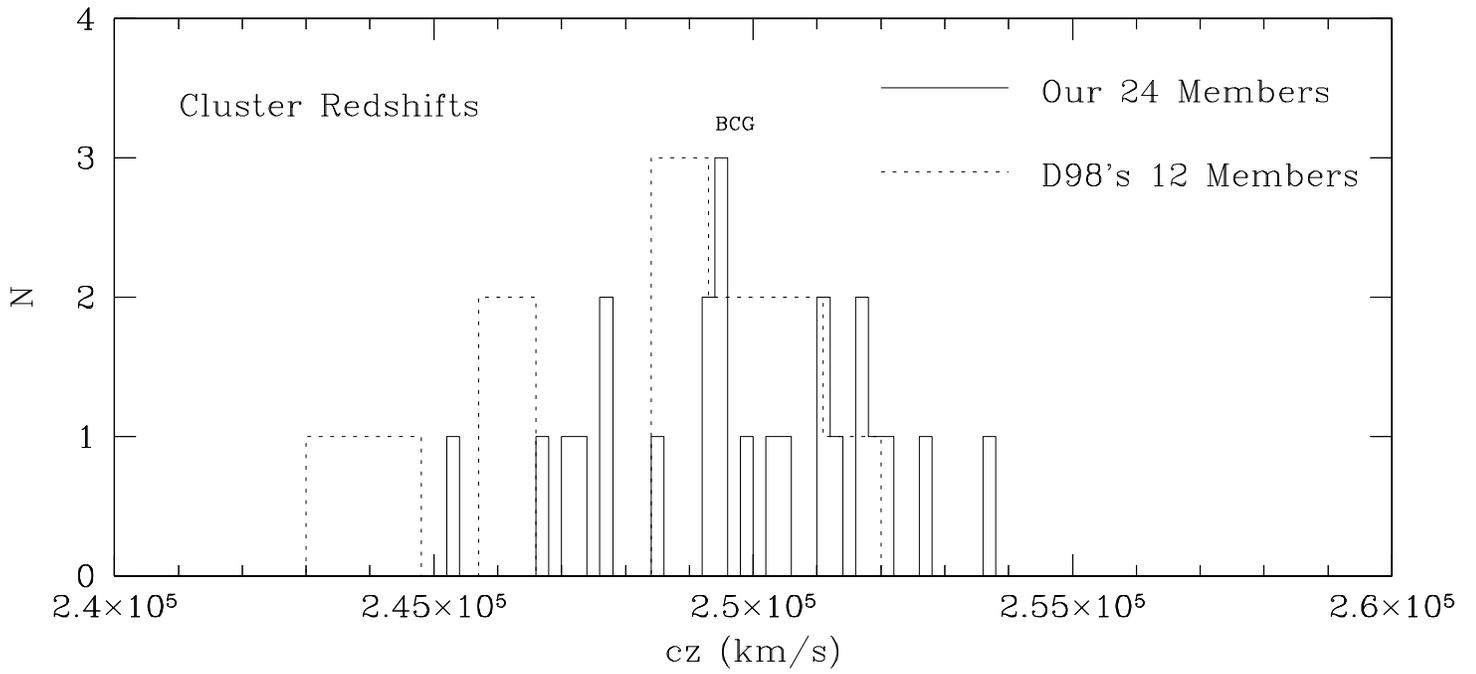
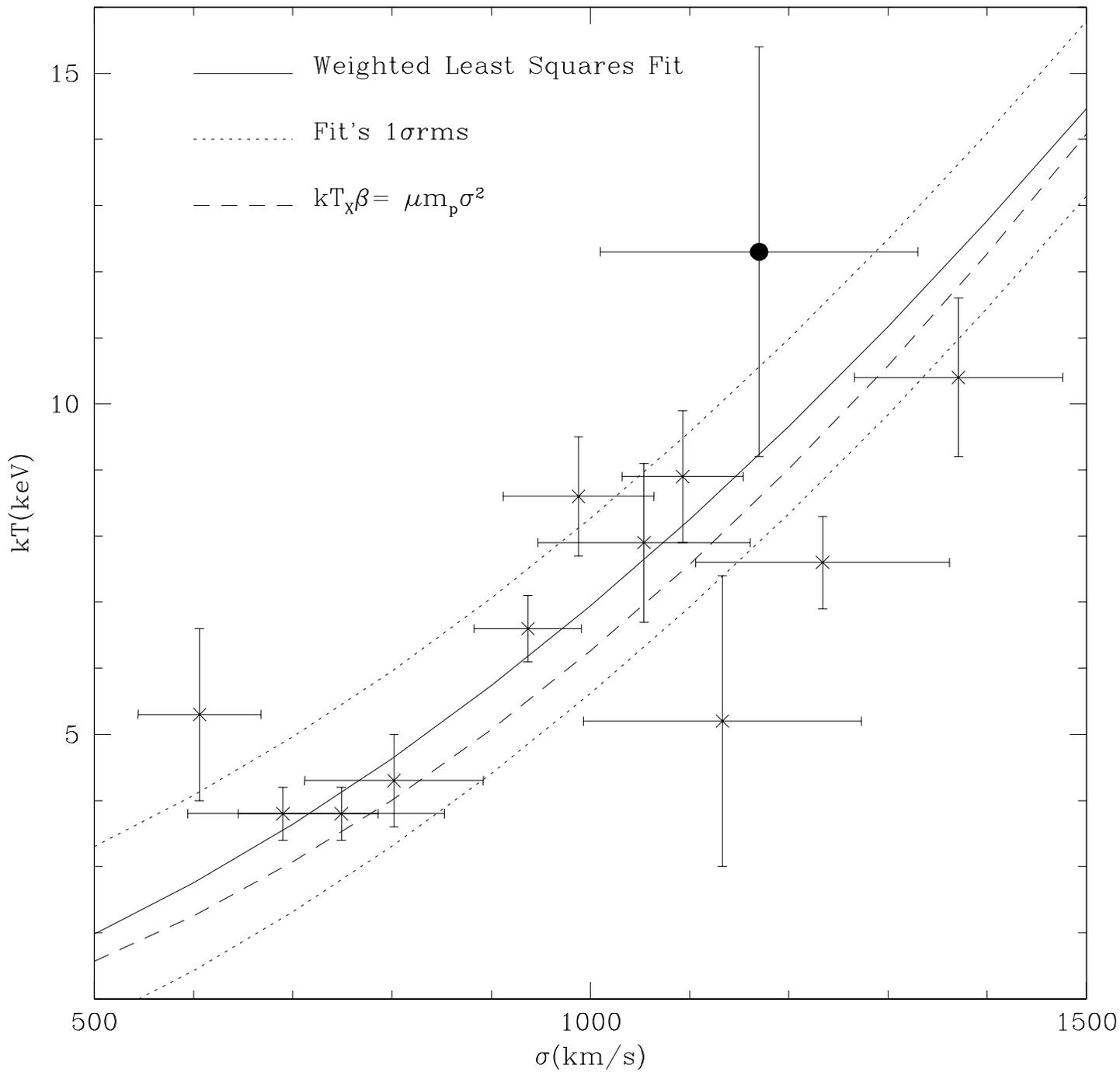
\end{document}